\documentclass[rnote]{aa}       

\usepackage{graphics}
\usepackage{epsfig}
\usepackage{latexsym}
\begin{document}
\hyphenation{brems-strah-lung}
\title{INTEGRAL and Swift/XRT observations of  IGR~J16460+0849}

\author{ Shu Zhang\inst{1}, Yupeng Chen\inst{1}, Jianmin Wang\inst{1,2},
        Tipei Li\inst{1,3}, Junqiang Ge\inst{1}
}

\institute{Laboratory for Particle Astrophysics, Institute of High
Energy Physics, Beijing 100049, China
\and
Theoretical Physics Center for Science Facilities (TPCSF), CAS
             \and
Center for Astrophysics,Tsinghua University, Beijing 100084, China
          }

\offprints{Shu Zhang}
\mail{szhang@mail.ihep.ac.cn}

\date{Received  / Accepted }

\titlerunning{INTEGRAL and Swift/XRT observations ...}
\authorrunning{Shu Zhang et al.}

  \abstract
  {}
{IGR J16460+0849 is reported in the 3rd IBIS catalog 
with the shortest exposure of about 10 ks among all the detected sources, which makes it the most interesting 
target to be investigated with a deeper exposure. The currently available data have increased to an exposure time of a few 
hundred ks. This stimulates us to investigate the source again  by
using all the available data. }
{We analyze all available observations  carried out by the International Gamma-Ray Astrophysics
Laboratory (INTEGRAL)  on the  unidentified source IGR J16460+0849.
 The data were processed by using the latest version OSA 7.0. 
In addition we  analyze also all the available Swift/XRT data on this source.}
{We find that IGR J16460+0849 has a detection significance of $\sim$ 4.4 $\sigma$ in the 20-100 keV band during the observational period between March 2003 and September 2004,
when it was exposed by $\sim$ 19 ks. Thereafter the source was not detected anymore, despite 
an additional exposure of $\sim$ 271 ks. This suggests a flux variability on a timescale of years. 
 The spectral analysis shows that the IBIS/ISGRI data are well presented by a power-law shape, with a photon index obtained as 1.45$\pm$0.86. So far, the source has been detected consistently by IBIS/ISGRI in the subsequent observations  and in the adjacent energy bands. 
We have analyzed the Swift/XRT observations on IGR J16460+0849 as well, and   found  no source inside the IBIS/ISGRI error circle. 
The non-detection  during the Swift/XRT observation is consistent with  the source having a variable nature.
} 
{} 

   \keywords{ X-rays: individual: IGR J16460+0849}

   \maketitle

\section{INTRODUCTION}

\begin{figure}[ptbptbptb]
\centering
 \includegraphics[angle=0, scale=0.25]{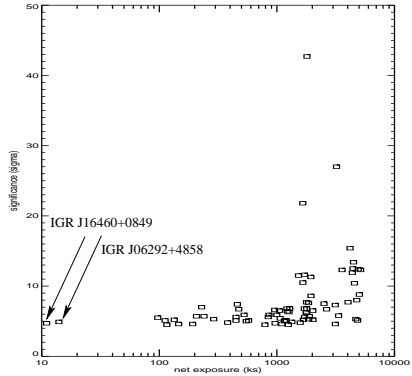}
      \caption{Source significance ($\sigma$) vs corrected onsource
exposure (ks) for sources listed in the 3rd IBIS catalog (Bird et al. 2007).}
         \label{sig-expo}
\end{figure}

Totally 421 sources are listed in the 3rd IBIS/ISGRI catalog 
(Bird et al. 2007). Among them are 171 Galactic accreting systems, 
122 extragalactic objects, and 113 of unknown nature. The diagram 
of {\it detection significance} versus {\it net exposure} for the unidentified 
sources is shown in Fig. \ref{sig-expo}. 
The net exposure refers to the corrected on-source exposure, e.g. omitting observations taken
during periods of solar activity or near the spacecraft perigee passages 
when the background modeling is difficult (Bird et al. 2007). Most sources have  
significances at the 5-7$\sigma$ level and are detected under  relatively 
large observational time of above hundreds  ks. In these cases, an additional exposure will not lead to a 
large increment in source significance.
However, the exceptions are IGR J06292+4858 and IGR J16460+0849. 
They are detected with a net exposure of only 14 ks and 11 ks, respectively. 
While for IGR J06292+4858 almost no additional exposure is available since Bird's report in 2007,
the situation has improved a lot for  IGR J16460+0849.  Therefore 
IGR J16460+0849 is the promising candidate for further analyses in the hard and soft X-ray bands, by using 
all currently available INTEGRAL and Swift/XRT data.

In this note we report the results from our INTEGRAL and Swift/XRT analyses on IGR J16460+0849, carried out by using the  most updated analysis tools. The paper is structured as follows: the observations and the 
data analyses are described in Sect.2, the results are presented in Sect.3, and finally 
in Sect.4 we give the discussion and the summary.

\section{OBSERVATIONS AND DATA ANALYSIS}
\subsection{INTEGRAL}

 INTEGRAL (Winkler et al. 2003) is a 15 keV - 10 MeV $\gamma$-ray mission. The main instruments are
 the  imager onboard INTEGRAL  (IBIS, 15 keV - 10 MeV) (Ubertini et al.
2003) and spectrometer 
(SPI, 20 keV - 8 MeV) (Vedrenne et al. 2003). They are supplemented by
the Joint European X-ray Monitor (JEM-X, 3-35 keV) (Lund et al. 2003) and the Optical
Monitor Camera (OMC, V, 500-600 nm) (Mas-Hesse et al. 2003).  At the lower energies (15 keV - 1 MeV), 
the CdTe array ISGRI (Lebrun et al. 2003) of IBIS has a better continuum sensitivity than SPI. 
The satellite was launched in October 2002
into an elliptical orbit with a period of 3 days. Due to the coded-mask
design of the detectors, the satellite normally operates in
dithering mode, which suppresses the systematic effects on spatial
and temporal backgrounds.

The INTEGRAL observations are carried out in the 
so-called individual  SCience Windows (SCWs), with a typical time duration of about 2000 seconds each.
 Only IBIS/ISGRI public data have been  taken
into account, while JEMX data are not available because the source was
outside its FOV.
The data reduction has been  performed by using the standard Online Science Analysis (OSA)
software version 7.0, the latest released version. The results 
are obtained by running the pipeline from the flowchart to the image level. 
The flux and the detection significance are measured from the mosaic map 
at the source position reported previously by Bird et al. (2007).

\subsection{Swift/XRT}

\emph{SWIFT} is a $\gamma$-ray burst explorer  launched 2004 November 20. 
It carries three co-aligned detectors (Gehrels et al. 2004), the Burst Alert Telescope 
(BAT, Barthelmy et al. 2005), the X-Ray Telescope (XRT, Burrows et al. 2005),  
and the Ultraviolet/Optical Telescope (UVOT, Roming et al. 2005). 
 We took only
Swift/XRT data into account, bacause BAT data were not available.
 The XRT uses a grazing incidence Wolter I telescope to focus X-rays onto 
a state-of-the-art CCD. XRT has an effective area of 110 cm$^2$, a FOV of 23.6 arcminutes, 
an angular resolution (half-power diameter) of 15 arcseconds, and it operates in the 0.2-10 keV energy range, providing 
the possibility of extending the investigation on the unidentified source to soft X-rays 
with the available SWIFT observations.

There is one SWIFT snapshot available for IGR J16460+0849 with an exposure of 4949 seconds. 
The observation was carried out on 2007 October 23 in the photon counting mode and has ID 00037052001. 
 We analyzed the Swift/XRT 0.2-10 keV data by using the latest released analysis software, provided in HEAsoft 
version 6.4. 

\section{RESULTS}
\subsection{INTEGRAL }

The available INTEGRAL observations, when IGR J16460+0849 falls into
 the partially coded field of view of ISGRI (offset angle less than 19
degrees),  comprise about 290
SCWs, adding up to a total exposure time of 511 ks (until 2006 August 25). 
 The details of the analyzed INTEGRAL observations on IGR J16460+0849, 
including the exposure and the time period, 
are summarized in Table 1. 
Most of these observations were carried out in the 5x5 dithering mode. We subdivided the data into 5 groups according to the 
observational sequence. For most of the time the source was at the edge of the field of view,
at an offset angle of more than 14 degrees.  In what follows the results are reported from the observations during which the source has the offset angle of less than 14 degrees.

 The imaging analyses show that the source is only detected  by summing the first three observational groups (MJD 52725-53259). 
It is not detected afterwards, although the exposure is increased by a factor of 14. The best source detection, $\sim$ 4.4 $\sigma$ at 20-100 keV 
(Fig. \ref{ima_16460_rev0056-0235}), is derived in the mosaic map of the first 3 observational groups.
The  flux in units of mcrab and the detection significance derived from each observational group in the 20-100 keV band are 
summarized in Table 1. We also generate the images in the 20-40 keV and 40-100 keV bands, with the combined  observations of the revolutions 0056-0235 (MJD 52725.0-53259.0). We derive the fluxes  of  7.4 mcrab and 12.4 mcrab
for the two energy bands, respectively.

\begin{figure}[ptbptbptb]
\centering
 \includegraphics[angle=0, scale=0.4]{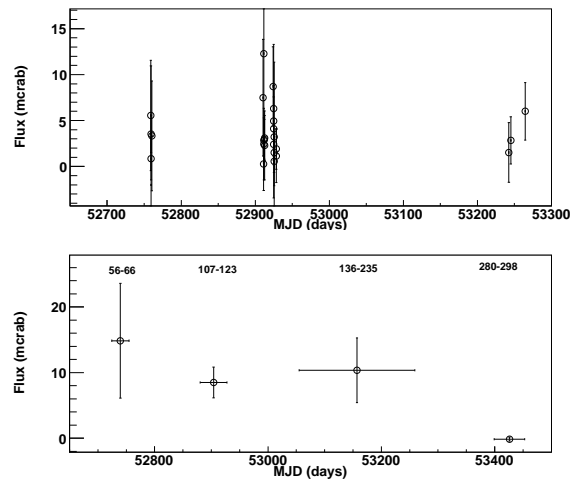}
      \caption{ISGRI longterm lightcurves of IGR J16460+0849 in the 20-100 keV band, on  the basis of SCW (top) 
and observational group (bottom), for the time period between 2003 and 2006. The revolution numbers are shown at the top of the bottom panel.}
         \label{lc_16460}
\end{figure}

 We produce in Fig. \ref{lc_16460} the SCW lightcurve for the observational groups containing the revolutions 0056-0235 (MJD 52725.0-53259.0), during which the source is observed to have the highest detection significance. There are only a few  SCWs with source offset angle of less than 14 degrees existing in each of the three observational groups. This prevents us from making a rough estimate of the duration of the source activity on a short time scale. To improve the statistics, we  add  the SCWs in each observational group together and produce the group lightcurve (Fig. \ref{lc_16460}). This lightcurve shows that the source stands out persistently  during the first three observational groups 0056-0235 (MJD 52725.0-53259.0), and does not become  detectable thereafter.  The average flux drops from roughly 9 mcrab of the first three groups to  an undetectable level of the later, clearly suggesting  a flux variability over a time scale of years.

\begin{figure}[ptbptbptb]
\centering
 \includegraphics[angle=0, scale=0.4]{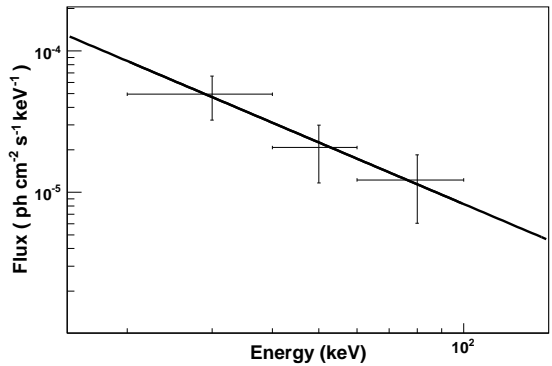}
      \caption{The energy spectrum of IGR J16460+0849 obtained from the observations of the revolutions 0056 - 0235 (MJD 52725.0-53259.0). The line shows the best fit with a power-law shape.  }
         \label{spectrum}
\end{figure}

The source spectrum is extracted from the observational sequence of the revolutions 0056-0235 (MJD 52725.0-53259.0) where the highest detection significance is presented.  The detection significances of the source in the energy bands  20-40 keV, 40-60 keV, and 60-100 keV are measured as 2.9 $\sigma$, 2.3 $\sigma$, and 2.0 $\sigma$, respectively. These data are well-fitted with a simple power-law model (see Fig. \ref{spectrum}), with the photon index  derived as  1.45$\pm$0.86 under a reduced $\chi^2$ of $\sim$ 0.1. The degree of freedom in the spectral fitting is 1.

\subsection{Swift/XRT}

 The Swift/XRT imaging analysis (5 ks exposure time) does not show any detected source inside the IBIS error circle (5 arcminutes radius).
We conclude that IGR J16460+0849 was not detected by Swift/XRT below 10 keV during this snapshot of 5 ks. Knowing the sensitivity of the XRT, we can put 
an upper  limit on the source flux. With an XRT sensitivity of about 2$\times$10$^{-14}$ erg cm$^{-2}$ s$^{-1}$ in 
an exposure time of 10 ks (see Capalbi et al. 2005), we estimate an upper flux limit of 
 $\sim$ 3$\times$10$^{-14}$ erg cm$^{-2}$ s$^{-1}$ from an observation of 5 ks.  
By extrapolating the ISGRI spectrum to energies below 10 keV, we have an integral flux of 
$\sim$ 1.3$\times$10$^{-10}$ erg cm$^{-2}$ s$^{-1}$ for the band 0.2-10 keV. 
 We would therefore  expect  a bright X-ray source from  Swift/XRT  exposure  of 5 ks if it was active at the time. The  non-detection  again shows that the source was variable and its flux dropped to a level below Swift/XRT threshold.

\section{Discussion and summary}

Bird et al. (2007) reported IGR J16460+0849 as an INTEGRAL hard X-ray source at a detection significance 
of 4.7 $\sigma$ in the 20-100 keV from a net exposure of $\sim$ 11 ks. Our analyses  agree with Bird's result, using the latest software release.  We find a source at a significance level of $\sim$ 4.4 $\sigma$ 
in the sum of the observations between March 2003 and September 2004. The source was not detected later, 
and the exposure increased to $\sim$ 271 ks. This suggests the source is most likely  variable on the timescale of years if the detected signal is real.

\begin{figure}[ptbptbptb]
\centering
\includegraphics[angle=0, scale=0.2]{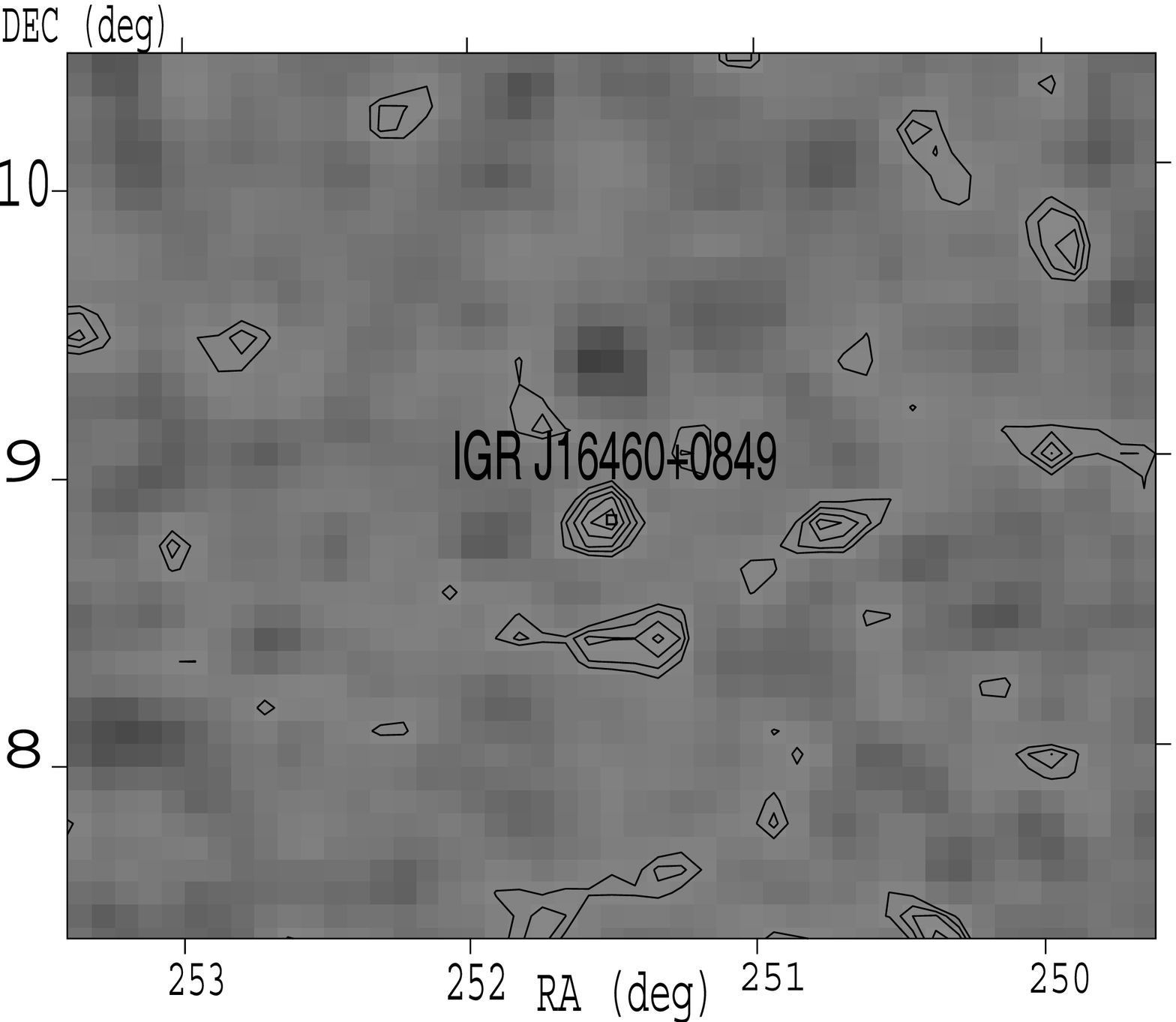}
 \includegraphics[angle=0, scale=0.2]{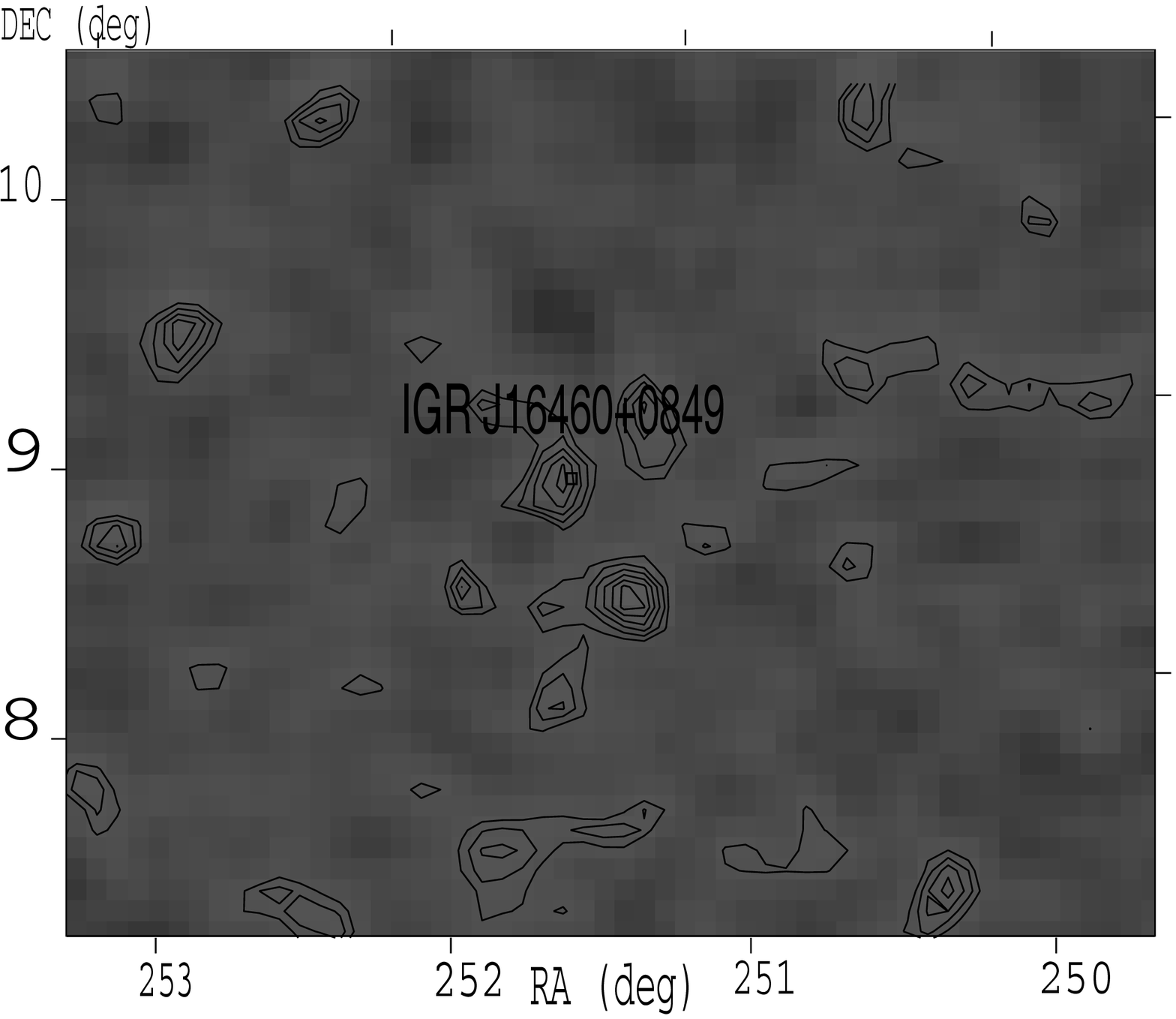}
      \caption{The 20-100 keV significance maps of IGR J16460+0849 as obtained 
from the combined  revolutions 0056 - 0235 (MJD 52725.0-53259.0).  The maps on the left and on the right  are  derived from data with a source offset angle less than 14 degrees and less than 19 degrees, respectively.  The contours start at a 
significance level of 2 $\sigma$ with steps of 0.5 $\sigma$.}
         \label{ima_16460_rev0056-0235}
\end{figure}

\begin{figure}[ptbptbptb]
\centering
 \includegraphics[angle=0, scale=0.4]{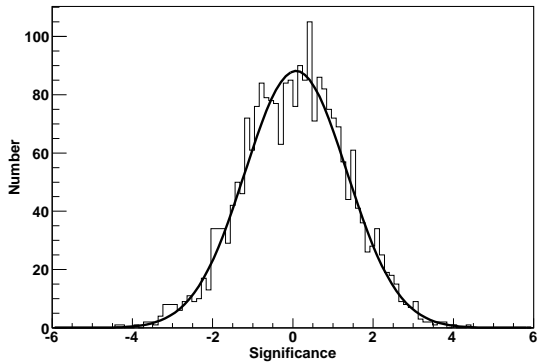}
      \caption{Gaussian fit to the significance distribution of the 20-100 keV mosaic map for 
 the time period of revolutions 0056 - 0235 (MJD 52725.0-53259.0).}
         \label{sig_gau}
\end{figure}

As listed in the 
3rd IBIS catalog (Brid et al. 2007), IGR J16460+0849 belongs to a group of  sources with low significances $<$ 5 $\sigma$. 
 About 10-20 percent of them might result from false detections (Bird et al. 2007). We notice that the map (Fig. \ref{ima_16460_rev0056-0235}), where the source has the  highest significance in detection, is rather noisy. 
To check whether this is statistical, we generated the histogram of the significance distribution. We find that the significance value for IGR J16460+0849 is consistent with the wing of a Gaussian shape, which  fits this distribution well (see Fig. \ref{sig_gau}). The 1 $\sigma$ variance of this distribution is about 1.27 and, accordingly, the probability of having a detection of 4.4 $\sigma$ decreases to 4.4/1.27$\sim$ 3.5 sigma under such a distribution. Nevertheless. we tend to believe the detected signal is more likely related to a real source, since even under such a low detection significance we still have  (1) the signals existing in adjacent three energy bands  which can be well fitted by a simple power law shape and, (2) the signals showing up in subsequent three observational groups over a time period of almost 1.5 years. Moreover, by comparing  the mosaic maps derived from the first three observational groups but with different  source offset angles  (14 degrees, Fig. \ref{ima_16460_rev0056-0235} left panel; 19 degrees,  Fig. \ref{ima_16460_rev0056-0235} right panel), we find that  IGR J16460+0849 stands  out  consistently  in both maps, while the noise structures can fluctuate considerably.

Bird et al. (2007) report a location of 
$(\alpha,\delta)_{2000}$ =  16h45m57s; +08d49m05s (l/b = 26.297$^{\circ}$/31.853$^{\circ}$) for IGR J16460+0849
with an error circle at the 90 percent confidence level of about  5 arcminutes. 
A SIMBAD search within this error location yields no known source. 
 The source location at middle Galactic latitude most probably indicates  an extragalactic origin, 
 so far for all AGNs reported in 3rd IBIS catalogue, additional exposure
leads to large increment in their significance, which clearly agrees
with their AGN nature. However, on the contrary, for IGR J16460+0849, additional 
IBIS exposure does not lead to an increment of its  significance but instead leads to a
significant decrease.  This agrees with the evidence that IGR J16460+
0849 is very likely a variable source spending a considerable fraction of the
time in a ``non active state", during which it is below the Swift/XRT and IBIS
threshold.

In summary, the most recent analyses carried out with the latest software releases     again show  
a hint of a source signal from  IGR J16460+0849. This is consistent with the previous report from Bird et al. (2007).   The high Galactic latitude of IGR
J16460+0849 suggests an extragalactic origin, but more data and further studies
are needed to shed light on its nature.

\begin{table}[ptbptbptb]
\begin{center}
\label{tab1}
\caption{ IBIS/ISGRI observations log of IGR J16460+0849. }
\vspace{5pt}
\small
\begin{tabular}{cccccc}
\hline \hline
Rev. &  Date  &  Expo.(${^\mathrm{a}}$)& Flux  &Sig.\\
 &  MJD  &  ks & mcrab &$\sigma$ \\
\hline
0056-0066 & 52725.0-52755.0 & 2.7(14) &14.8$\pm$8.7 &1.7\\
\hline
0107-0123 & 52880.5-52927.5 & 11.7(52)  &8.5$\pm$2.4&3.6\\
\hline
0136-0235 & 53055.0-53259.0 & 5.0(40) &10.3$\pm$5.0&2.1\\
\hline
0280-0298 & 53399.1-53453.0 & 271(272)&0&0\\
\hline
0345-0472($^{\mathrm{b}}$) & 53592.7-53972.3 & 0(133) &--&--\\
\hline
0056-0235 & 52725.0-53259.0 &  20(106) &9.0$\pm$2.0 &4.4\\
0056-0472 & 52725.0-53972.3 &  300(511) &0.04$\pm$0.2 &0.2\\
\hline
\hline
\end{tabular}
\end{center}
\begin{list}{}{}
\item[${\mathrm{a}}$]{the first value is the exposure with a source offset angle less than 14 degrees, and the value in parentheses 
is the exposure with a source offset angle less than 19 degrees.}
\item[$^{\mathrm{b}}$]{ the source is never at an offset
angle less than 14 degrees.}
\item[Note:]{The flux and the significance are presented in the energy band 20-100 keV for the data where the source was within an offset angle of  14 degrees.}

\end{list}
\end{table}

\acknowledgements
 We thank the anonymous referee and Dr. Werner Collmar for the constructive comments that   helped in improving our paper. 
 This work was subsidized by the National Natural Science Foundation of China,  the CAS key Project KJCX2-YW-T03, and the 973 Program 2009CB824800. J.-M. W. thanks the Natural Science Foundation of China for support via NSFC-10325313, 10521001, and 10733010.

\end{document}